\documentclass[english,aps,prl,superscriptaddress,reprint,longbibliograhy]{revtex4-2}
\usepackage[T1]{fontenc}
\usepackage[utf8]{inputenc}
\usepackage{lmodern}
\usepackage{amsmath,amssymb,mathtools,bm}
\usepackage{graphicx}
\usepackage{siunitx} 
\usepackage[colorlinks=true,allcolors=blue]{hyperref}
\usepackage{color}
\definecolor{darkgreen}{RGB}{0,100,0} 
\setcitestyle{numbers,sort&compress} 

\begin{document}

\title{Universal theory of domain-wall width in multi-sublattice Heisenberg magnets}

\author{Jose M. Lendínez}
  \affiliation{Instituto de Ciencia de Materiales de Madrid, CSIC, Cantoblanco, 28049 Madrid, Spain\looseness=-1}

\author{Marta Yanguas}
  \affiliation{Instituto de Ciencia de Materiales de Madrid, CSIC, Cantoblanco, 28049 Madrid, Spain\looseness=-1}

\author{Theodor Griepe}
  \affiliation{Instituto de Ciencia de Materiales de Madrid, CSIC, Cantoblanco, 28049 Madrid, Spain\looseness=-1}

\author{Michael Saur}
  \affiliation{Instituto de Ciencia de Materiales de Madrid, CSIC, Cantoblanco, 28049 Madrid, Spain\looseness=-1}

\author{Rubén M. Otxoa}
  \affiliation{Hitachi Cambridge Laboratory, J. J. Thomson Avenue, Cambridge CB3 0HE, United Kingdom\looseness=-1}
  
\author{Levente Rózsa}
   \affiliation{Department of Theoretical Solid State Physics, Institute of Solid State Physics and Optics,
HUN-REN Wigner Research Centre for Physics, H-1525 Budapest, Hungary}
   \affiliation{Department of Theoretical Physics, Budapest University of Technology and Economics, M\H{u}egyetem rkp. 3, H-1111 Budapest, Hungary}
    
\author{Unai Atxitia}
  \affiliation{Instituto de Ciencia de Materiales de Madrid, CSIC, Cantoblanco, 28049 Madrid, Spain\looseness=-1}

\date{\today}

\begin{abstract}
We propose a universal expression for the domain-wall width in generic multi-sublattice Heisenberg magnets, applicable to ferro-, antiferro-, and ferrimagnetic orders. The result follows from an exact connection between the domain-wall profile and the long-wavelength spin-wave dispersion, yielding a unified framework for describing magnetic textures across distinct ordering types. 
The predictions show excellent quantitative agreement with large-scale atomistic spin dynamics simulations over a broad range of exchange and anisotropy values and spin multi-sublattice structures, including three-dimensional rock-salt-type magnets and two-dimensional honeycomb and kagome ferromagnets. Moreover, we establish a general microscopic foundation for the temperature dependence of the domain-wall width. Our approach offers a powerful tool for understanding domain-wall profiles in complex spin systems.
\end{abstract}

\maketitle

Magnetic domain-walls (DWs) are prototypical topological textures whose structure and dynamics play a central role across condensed-matter physics, magnetism, and modern spintronics \cite{Parkin2008,Caretta2020,Wang2025}. They underpin a wide range of emergent phenomena, from chirality-dependent transport \cite{Emori2013} to ultrafast angular-momentum flow \cite{Hennecke2025}. In spintronic technologies, DWs act as mobile information carriers in racetrack-memory architectures \cite{Parkin2015} and logic devices \cite{Luo2020}. They can be efficiently driven by spin–orbit torques \cite{Manchon2019}, acoustic waves \cite{Rivelles2025}, and thermal gradients \cite{Schlickeiser2014,Donges2020}, and enable sub-nanosecond control when engineered in ferri- or antiferromagnetic materials \cite{Hardt2025,Diona2025}.

The DW width is a fundamental parameter linking microscopic exchange and anisotropy to macroscopic device functionality \cite{kumar_domain_2022}. By setting the length scale of magnetization rotation, it determines the energy, stability, inertia, mobility, and dynamic response of DWs under external stimuli \cite{HubertSchaefer1998, Malozemoff1979Bubble, schryer_motion_1974, wu_high-mobility_2026}. It also governs magnetoresistance by controlling the adiabaticity of electron spin transport and affects the efficiency of spin-transfer and spin-orbit torques \cite{li_domain-wall_2004}. Despite its central role, a complete understanding on the parameters controlling the DW width is still lacking. The compact expression $\delta_{\rm DW}=\sqrt{A/K}$, where $A$ is the exchange stiffness and $K$ the uniaxial anisotropy, accurately characterizes the DW width in simple ferro- and antiferromagnets \cite{Aharoni2000Ferromagnetism, theodorou_antiferromagnetic_domain}.  This description, however, breaks down in complex systems with multiple sublattices and competing energy scales \cite{Moreno2025}. 


In this Letter, we propose a universal  expression for the DW width in multi-sublattice Heisenberg magnets with uniaxial anisotropy. This relation emerges from an explicit connection between the spin-wave (SW) dispersion and the effective parameters that determine the DW width, thereby establishing a unified framework across different magnetic orders. Our predictions show excellent quantitative agreement with large-scale atomistic spin dynamics  (ASD) simulations  \cite{Nowak2007ClassSpinModels,Evans_2014_JPhysCondensMatter_26_10}. Atomistic simulations further enable the incorporation of finite-temperature effects through the thermal renormalization of spin-model parameters \cite{bottcher_temperature-dependent_2012, szilva_interatomic_2013}, yielding a self-consistent description of DWs from zero temperature up to the magnetic ordering transition. 

 In ferromagnets, both the exchange stiffness and the anisotropy that determine $\delta_{\rm DW}$ are encoded in the long-wavelength limit of the SW spectrum. This reflects a deep correspondence: SWs are small-amplitude propagating twists of the magnetic order, whereas a DW is a static, spatially localized twist of the same order parameter \cite{coey_magnetism_2010}. Formally, the curvature of the SW dispersion near $q \to 0$ defines the effective stiffness, $2A \propto \partial_q^2 \varepsilon_q \big|_{q=0}$, while the energy gap 
$2K\propto \varepsilon_0$ encodes the magnetic anisotropy \cite{lifshitz_theory_1951, herring_theory_1951}. The competition between the exchange stiffness suppressing spin modulations and the anisotropy aligning the spins along a preferred direction yields the expression for the domain-wall width, $\delta_{\rm DW}=\sqrt{A/K}$. However, in complex multi-sublattice  spin systems, such as antiferromagnets, ferrimagnets or 2D ferromagnets, the situation is more subtle: multiple SW branches, sublattice-dependent exchange couplings and mode hybridization obscure the identification of the relevant parameters determining the DW width \cite{macneill_gigahertz_2019, xu_inversion_2026, chernyshev_damped_2016}. 
Consequently, no general relation exists that directly links the measurable SW spectrum to the DW width in multi-sublattice spin systems.

\begin{figure}[t]
\includegraphics[width=\linewidth]{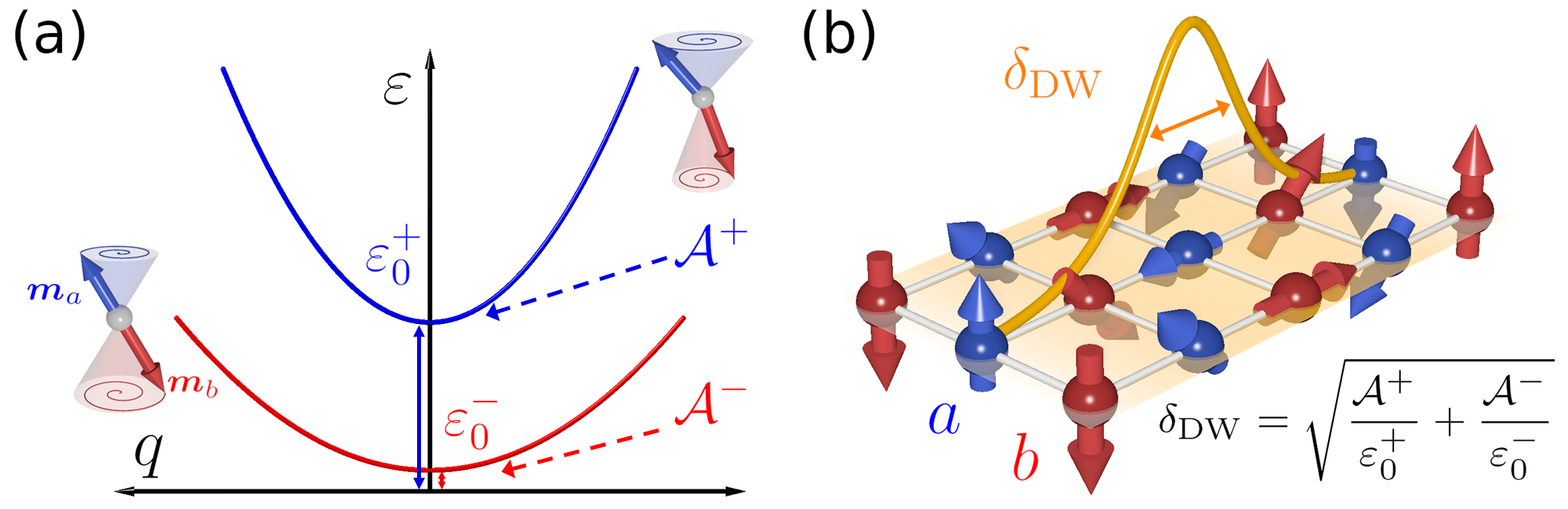}

\caption{(a) Schematic representation of the spin-wave spectrum of a two-sublattice ferrimagnet, characterized by energy modes $\varepsilon_q^{\pm}$ with gaps $\varepsilon_0^{\pm}$ and spin-wave stiffnesses $\mathcal{A}^{\pm} \propto \partial_q^2 \varepsilon^{\pm}_q \big|_{q=0}$.
(b) Sketch of a domain-wall of  width $\delta_{\rm DW}$ in a two-sublattice ferrimagnet. The orange line indicates the transversal component of the sublattice magnetization.
}
\label{fig:fig1}
\end{figure}

Our work establishes a universal connection, showing that the DW width can be written in the compact form
\begin{equation}
\label{eq:UniversalDW}
\delta_{\rm DW}=\sqrt{\sum_\beta \mathcal{A}^{\beta}/\varepsilon^{\beta}_0},
\end{equation}
where the spin-wave spectrum is approximated as
\begin{equation}
\varepsilon_q^{\beta}=\varepsilon_0^{\beta }+\mathcal{A}^{\beta}q^2, 
\label{eq:dispersion_beta}
\end{equation} with $\mathcal{A}^{\beta}$ denoting the SW stiffness and $\varepsilon_0^{\beta}$ the energy gap of  $\varepsilon_q^\beta$ mode, as schematically illustrated in Fig.~\ref{fig:fig1} for a two-sublattice ferrimagnet. As we demonstrate below using large-scale atomistic spin dynamics simulations, this expression provides a direct link between microscopic SW properties and the continuum description of DWs in uniaxial Heisenberg magnets. 

\begin{figure}[t]
\centering
\includegraphics[width=1.0\linewidth]{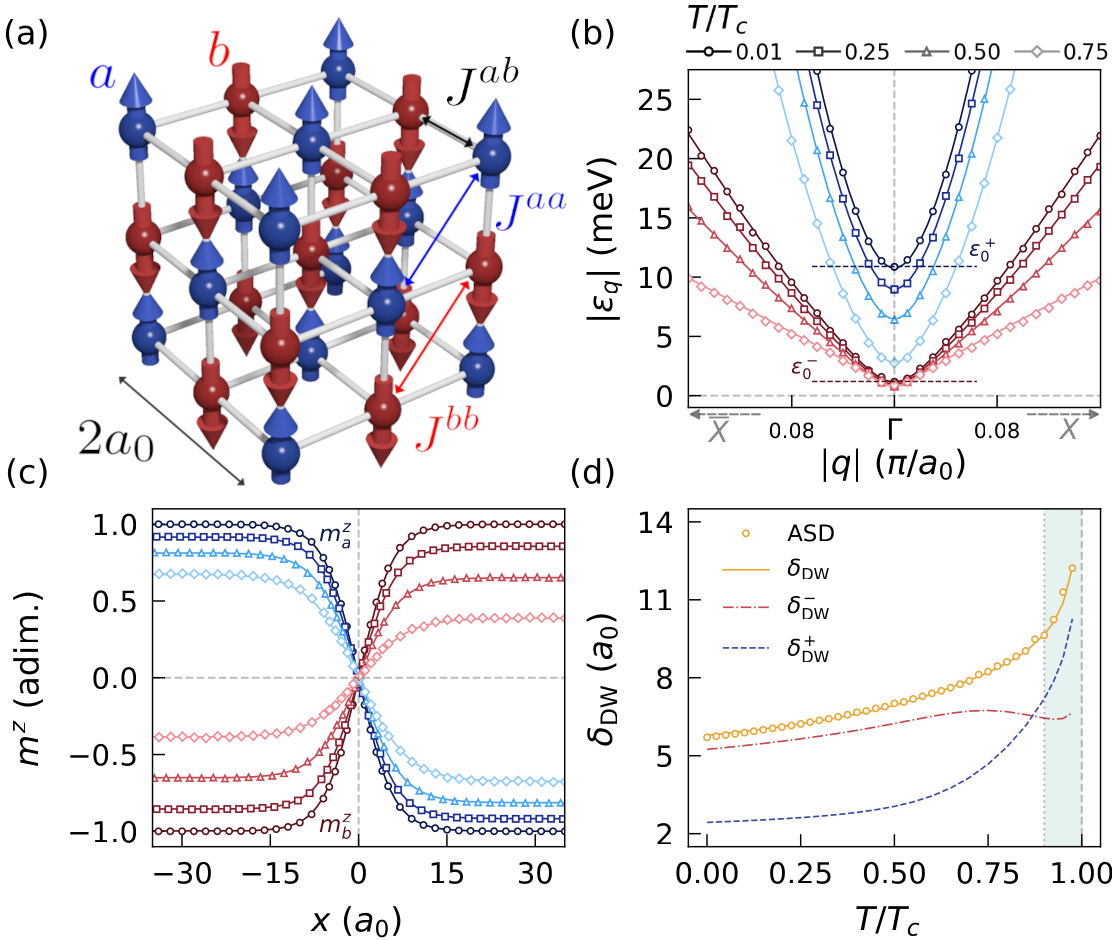}
\caption{(a) Sketch of the rock-salt–type magnet used in the atomistic spin model, illustrating ferrimagnetic (FI) ordering. (b) Spin-wave spectrum of the FI system near $\Gamma$ along the $\overline{\Gamma X}$ direction ($\overline{\Gamma X}=\pi/a_0$) at selected reduced temperatures $T/T_c$. (c) Reduced sublattice magnetization $m^z_r$ profiles at $T/T_c$. (d) DW width as a function of  $T/T_c$.  Symbols correspond to ASD simulations in panels (b-d). Solid lines indicate fits to the SW dispersion in (b) and the universal-relation values from  Eq.~\eqref{eq:UniversalDW} in (c,d). The green-shaded area in (d) marks the region above the compensation temperature of the FI system.}
\label{fig:fig2}
\end{figure}

As a first example, we consider a rock-salt-type, two-sublattice [see Fig.~\ref{fig:fig2}(a)] uniaxial Heisenberg model, described by the classical atomistic  spin Hamiltonian
\begin{equation}
    \mathcal{H}= -\frac{1}{2}\sum_{i,j,r,s} J^{rs}_{ij}\mathbf{S}_{ir}\mathbf{S}_{js} - \sum_{i,r}D_r(S_{ir}^z)^2, 
    \label{eq:Hamiltonian}
\end{equation}
where $i$ and $j$ 
label lattice sites and $r,s \in \{a, b\}$ denote the two magnetic sublattices. The variables $\mathbf{S}_{ir}$ are classical spins of fixed length $|\mathbf{S}_{ir}|=1$ and the magnetic moments are given by $\boldsymbol{\mu}_{ir}= \mu_r \mathbf{S}_{ir}$. The first term describes isotropic Heisenberg exchange interactions, where $J_{ij}^{rs}$ is the exchange coupling between spins on sites $i$ and $j$ belonging to sublattices $r$ and $s$, respectively. 
We only consider nearest-neighbor couplings $J^{ab}$ between the sublattices and next-nearest-neighbor couplings, $J^{aa}$ and $J^{bb}$, inside each sublattice. The second term represents a single-ion uniaxial magnetocrystalline anisotropy $D_r$ on sublattice $r$.
Depending on the choice of the coupling parameters, this Hamiltonian can describe a ferrimagnet ($J^{ab}<0$ and $\mu_a \neq \mu_b$), a ferromagnet on a face-centered cubic lattice ($J^{ab}=J^{bb}=0$) or a simple cubic antiferromagnet ($J^{ab}<0$, $J^{aa}=J^{bb}=0$ and $\mu_a = \mu_b$).

The SW spectrum of a two-sublattice magnet consists of two branches with energies $\varepsilon^{\pm}_{q}$, determined by the parameters defining the Hamiltonian in Eq.~\eqref{eq:Hamiltonian}. Since our Hamiltonian contains isotropic exchange interactions, in the long-wavelength limit the full SW dispersion reduces to the well-known quadratic form in Eq.~\eqref{eq:dispersion_beta} for $\beta=\pm$.
The SW energy gaps, $\varepsilon_0^{\pm}$, and stiffnesses, $\mathcal{A}^{\pm}$, can be derived analytically within 
linear spin-wave theory (LSWT). 
We note that for a single-sublattice ferromagnet the SW dispersion reduces to one branch, \mbox{$\varepsilon_{q}=\varepsilon_0+\mathcal{A}q^2$}. Within LSWT, this already provides the information needed to predict the DW width at zero temperature [see the Supplemental Material (SM)~\cite{SM} for details]. At finite temperature, however, thermal populations of SW modes renormalize the relevant parameters \cite{Bastardis2012, Rozsa2023}. To account for this temperature-dependent renormalization, we compute the SW spectra using ASD simulations, incorporating thermal fluctuations up to the critical temperature. The resulting spectra yield the renormalized exchange and anisotropy constants, as well as the thermally averaged magnetic moments entering the LSWT expressions. These relations allow the calculation of the temperature dependence of $\delta_{\rm DW}$ via Eq.~\eqref{eq:UniversalDW}.

We validate our results through ASD simulations of stationary DW profiles. The DW width is extracted by fitting the sublattice-resolved magnetization profiles along the spatial coordinate $x$ (defined as the direction perpendicular to the DW plane) to
\begin{equation}
\begin{aligned}
&m^{z}_{r}(x,T) = m_r(T)\,\tanh\!\left[\left(x - x_0\right)/\delta_{\mathrm{DW}}^{\mathrm{ASD}}(T)\right], \\
&m^{\perp}_{r}(x,T) = \frac{m_r(T)}
{\cosh \left[ \left(x - x_0\right)\,/\, \delta_{\mathrm{DW}}^{\mathrm{ASD}}(T)\right]}.
\end{aligned}
\label{eq:dw_profile}
\end{equation}
where $m^z_r$, $m^{\perp}_r$, and $m_r$ denote the longitudinal component, transverse component, and saturation magnetization of sublattice $r$ at temperature $T$, respectively. The DWs are centered at $x_0$, while $\delta_{\mathrm{DW}}^{\mathrm{ASD}}(T)$ is obtained from the fit and defines the corresponding DW width [see SM~\cite{SM} for computational details].

To demonstrate the universality of our approach, we first consider a generic ferrimagnetic system (FI) in Fig.~\ref{fig:fig2}. Our atomistic spin model consists of two inequivalent sublattices with uncompensated magnetic moments, along with the exchange interactions and anisotropy parameters given under FI in Table~\ref{tab:parameters}.

\begin{table}[t]
\caption{Microscopic material parameters for the rock-salt-type magnets studied here}
\label{tab:parameters}
\centering
\renewcommand{\arraystretch}{1.25} 
\resizebox{0.9\columnwidth}{!}{
\begin{tabular}{ccccccc}
\hline\hline
System & $J^{aa}$ & $J^{bb}$ & $J^{ab}$ &
$\mu_a$ & $\mu_b$ & $D_a=D_b$ \\
& (meV) & (meV) & (meV) &
($\mu_{\rm B}$) & ($\mu_{\rm B}$) & (meV) \\
\hline
FI  & 10 & 5 & -2.5 & 2 & 5 & 0.5 \\
FM  & 10 & 0 & 0 & 2 & 0 & 0.05 ($D_a)$\\
AFM & 0 & 0 & -20 & 2 & 2 & 0.05 \\
\hline\hline
\end{tabular}
}
\end{table}

We first discuss the agreement in the zero-temperature limit, approximated here by $T/T_c = 0.01$, where $T_c$ stands for the critical temperature for the corresponding order parameter. The SW spectrum consists of two modes: a high-energy branch $\varepsilon^{+}_q$ and a low-energy branch $\varepsilon^{-}_q$, each with distinct gaps and SW stiffness values [see Fig.~\ref{fig:fig2}(b)]. The analytical expressions for $\varepsilon_0^{\pm}$ and $\mathcal{A}^{\pm}$ parameters are detailed in the SM \cite{SM}. From Eq. \eqref{eq:UniversalDW}, we derive a simplified expression for $\delta_{\rm DW}$ within the exchange dominated (Exch.-D.) regime, i.e., when the anisotropy is relatively small compared to the exchange energies [see SM~ \cite{SM}], 
\begin{equation}
\delta_{\rm{DW}}^{\mathrm{Exch.}\text{-}\rm{D.}}\approx\sqrt{\frac{\mathcal{A}^+}{\varepsilon_0^+}+\frac{\mathcal{A}^-}{\varepsilon_0^-}} \approx a_0\sqrt{  \frac{2J^{aa}+2J^{bb}-J^{ab}}{(D_a+D_b)}}.
\label{eq:exchange_dom}
\end{equation}
Equation~\eqref{eq:exchange_dom} can be viewed as a ferrimagnetic generalization of the standard ferromagnetic relation \mbox{$\delta_{\rm DW}=\sqrt{A/K}$}, with an effective exchange stiffness given by  $A/2=(2J^{aa}+2J^{bb}-J^{ab})/a_0$ and a macroscopic anisotropy density   $K/2=(D_a+D_b)/a^3_0$~\cite{Moreno2025}.



At finite temperatures, thermal fluctuations renormalize the material-specific parameters, as reflected in the temperature-dependence of the SW spectrum [see Fig.~\ref{fig:fig2}(b)]. By extracting the renormalized parameters from fits to the LSWT expression, Eq.~\eqref{eq:UniversalDW} captures the temperature dependence of the DW width up to $T_c$ [see Fig.~\ref{fig:fig2}(c)] and directly relates its evolution to the changes in the two SW branches. Figure~\ref{fig:fig2}(d) illustrates the temperature dependence of $\delta_{\rm DW}$ and the partial contributions \begin{equation}
    \delta_{\rm DW}^{\pm}=\sqrt{\mathcal{A}^{\pm}/\varepsilon_0^{\pm}}.
\end{equation} At low temperatures, the DW width is primarily determined by $\delta_{\rm DW}^{-}$, with 
a small but noticeable correction from the high-energy branch $\varepsilon_q^{+}$.
As temperature increases, both gaps decrease and $\delta_{\rm DW}$ grows, while the high-energy contribution remains minor.
Notably, near the compensation point, the energy gap between branches vanishes, $\varepsilon_0^{+} - \varepsilon_0^{-} \rightarrow 0$~\cite{barker_two-magnon_2013}. As a direct consequence, $\delta_{\rm DW}^{+}$ becomes comparable to $\delta_{\rm DW}^{-}$. Above the compensation point, the $\varepsilon_q^{+}$ branch becomes the lower-energy branch and dominates the overall behavior of the DW width.


\begin{figure}[t]
\centering
\includegraphics[width=1.0\linewidth]{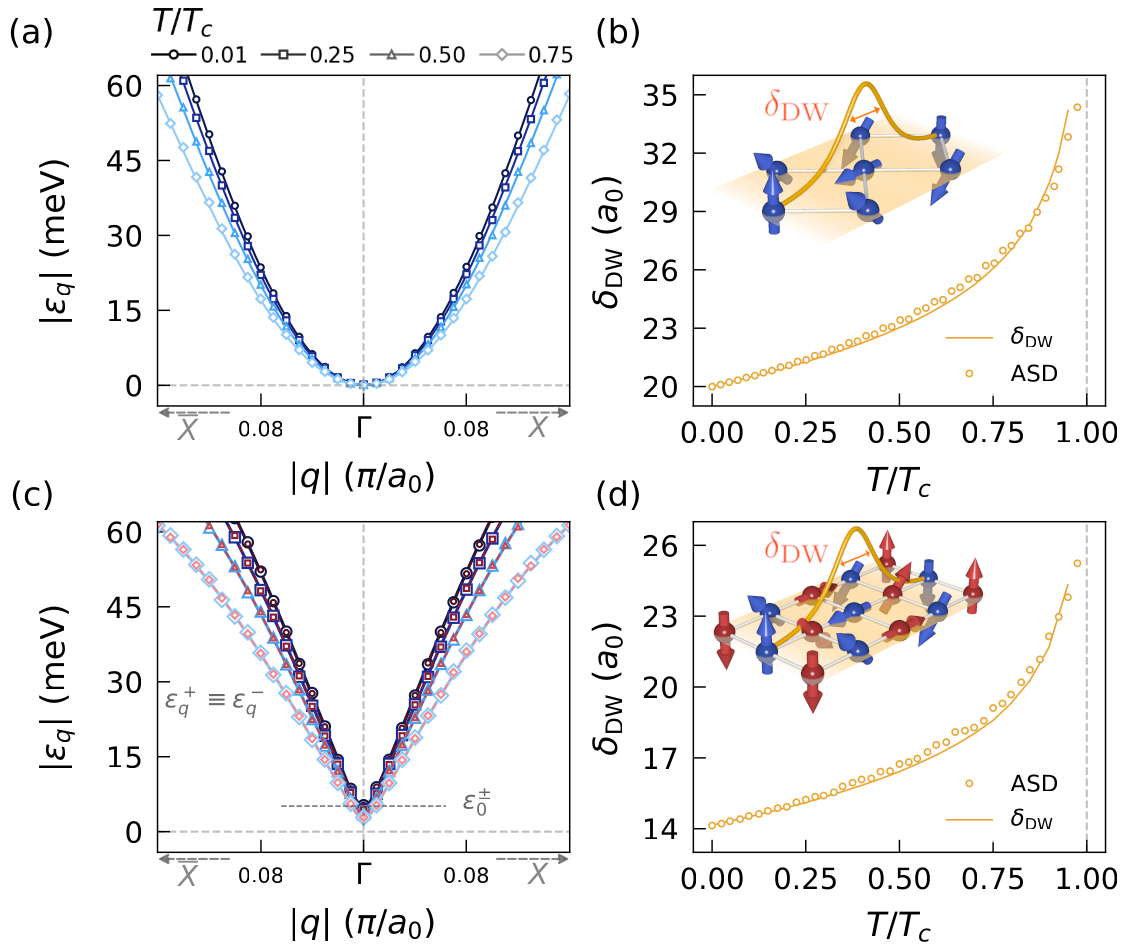}
\caption{Face-centered cubic ferromagnet: (a) Spin-wave spectrum near $\Gamma$ along $\overline{\Gamma X}$ direction at selected $T/T_c$. (b) DW width as a function of $T/T_c$. Simple cubic antiferromagnet: (c) Spin-wave spectrum near $\Gamma$ along $\overline{\Gamma X}$ direction at selected $T/T_c$ (b) DW width as a function of $T/T_c$. Symbols correspond to ASD simulations in all panels. Solid lines in (a,c) and in (b,d) indicate fits to the SW dispersion and values from the universal relation in Eq.~\eqref{eq:UniversalDW}, respectively. Panels (b, d) include a sketch of the DW profile for a slice of the system.}
\label{fig:fig3}
\end{figure}
To demonstrate the generality of our approach, we show that it applies to both ferromagnetic and antiferromagnetic orderings in the rock-salt structure.

A single-sublattice ferromagnet (FM) on a rock-salt lattice reduces to a face-centered cubic structure, characterized here by nearest-neighbor intrasublattice exchange and uniaxial anistropy given under FM in Table.~\ref{tab:parameters}. The SW spectrum obtained from ASD simulations [see Fig.~\ref{fig:fig3}(a)] consists of a single branch with energy gap $\varepsilon_0 = 2D_a$ and stiffness $\mathcal{A} = 4J^{aa} a_0^2$. These expressions yield a DW width of $\delta_{\rm DW} = a_0 \sqrt{2J^{aa}/D_{a}} = 20\,a_0$ at zero temperature, in excellent agreement with direct ASD calculations [see Fig.~\ref{fig:fig3}(b)]. 

An antiferromagnet on a rock-salt lattice reduces to a simple cubic structure. The microscopic material parameters of our AFM system are detailed in Table.~\ref{tab:parameters}. In contrast to the FM, the SW spectrum of an AFM [see Fig.~\ref{fig:fig3}(c)] consists of two degenerate branches with both exchange-enhanced energy gaps $|\varepsilon_0^{\pm}| =2\sqrt{-6DJ^{ab}}$ and SW stiffnesses $|\mathcal{A}^{\pm}|= 3(J^{ab})^2a_0^2/\sqrt{-6DJ^{ab}}$.  This case highlights the importance of considering the contribution from both branches for an exact calculation of the DW width [see Fig.~\ref{fig:fig3}(d)], yielding \mbox{$\delta_{\rm DW}=a_0\sqrt{-J^{ab}/4D-J^{ab}/4D}=a_0\sqrt{-J^{ab}/2D}$}. As in the FM case, our approach also captures the temperature dependence of the AFM domain-wall width [see Fig.~\ref{fig:fig3}(d)].

\begin{figure}[t]
\centering   
\includegraphics[width=0.465\textwidth]{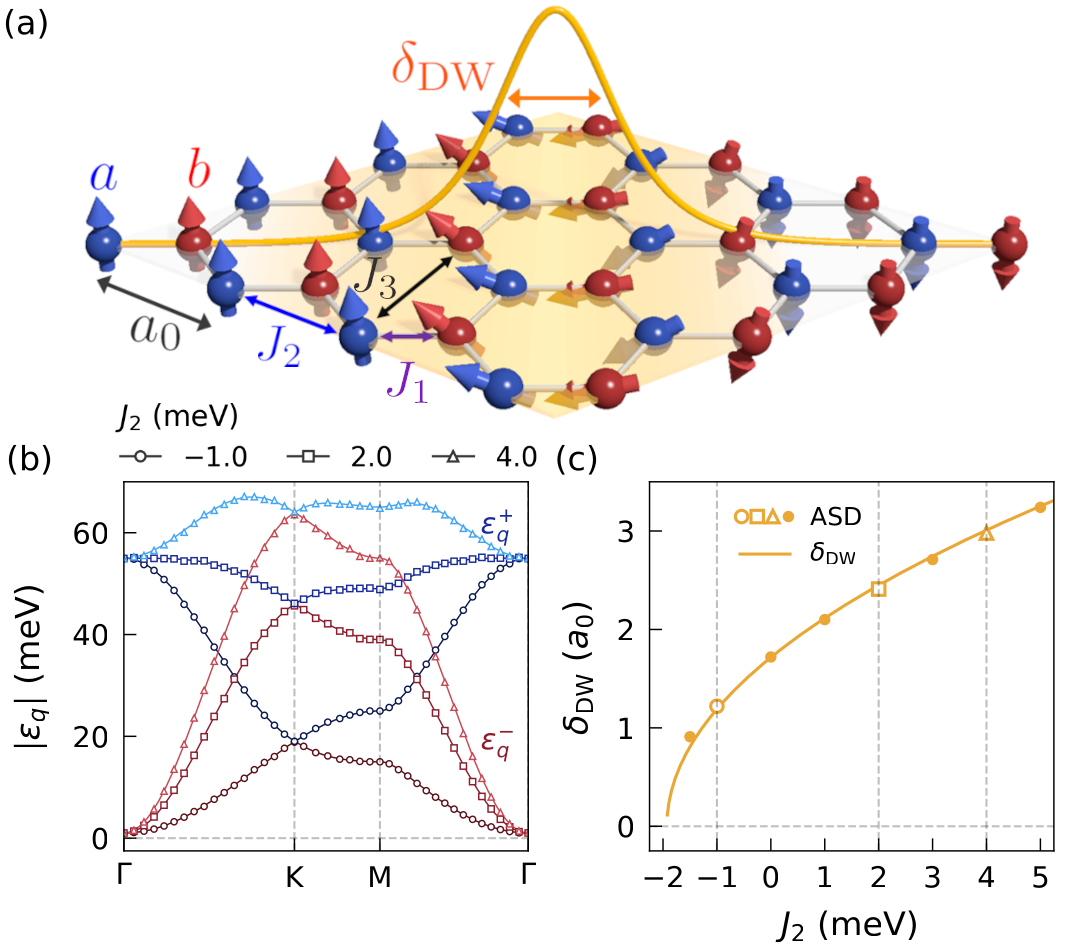}\caption{ (a) Sketch of a DW profile in a 2D honeycomb ferromagnet.
(b) Spin-wave spectrum along the indicated high-symmetry path at zero temperature for selected values of $J_2$.
(c) DW width as a function of $J_2$. Solid lines denote LSWT and analytical results for panels (b) and (c), respectively. Symbols correspond to ASD simulations; open symbols in (c) correspond to the selected cases shown in (b).}
    \label{fig:fig4}
\end{figure}

As another example of the universality of Eq.~\eqref{eq:UniversalDW}, we consider a two-dimensional (2D) honeycomb lattice, a bipartite structure composed of two interpenetrating triangular sublattices. Depending on the exchange couplings and sublattice moments, the lattice can exhibit ferromagnetic, antiferromagnetic, or ferrimagnetic ground states. Here, we focus on the exchange interactions $J_1$, $J_2$, and $J_3$, corresponding to first-, second-, and third-nearest neighbors as depicted in Fig.~\ref{fig:fig4}(a). We consider identical sublattice-specific parameters that yield ferromagnetic ordering and include a uniaxial on-site anisotropy $D$ [see Fig.~\ref{fig:fig4}(a)]. Uniaxial anisotropy stabilizes long-range order, circumventing the Mermin–Wagner theorem \cite{Hohenberg1967,Mermin1966,Jenkins2022} and making honeycomb lattices relevant for atomically thin magnets such as CrI$_3$, CrBr$_3$, and CrGeTe$_3$ \cite{Cenker2021,Chen2018}. Our study is limited to zero temperature due to enhanced fluctuations associated with reduced dimensionality.

Remarkably, we show that the universal relation in Eq.~\eqref{eq:UniversalDW} holds for the honeycomb ferromagnet, acquiring the simple form
\begin{equation}
    \delta_{\rm{DW}}= \sqrt{\frac{6J_2 + \left(J_1 + 4J_3 \right)}{8D}+\frac{6J_2 - \left( J_1 + 4J_3 \right)}{8D+24(J_1+J_3)}},
    \label{eq:honeycombA+}
\end{equation}
which arises from the contributions of both SW branches, with the first term under the square root corresponding to  $\mathcal{A}^{-}/\varepsilon_0^{-}$, and the second to $\mathcal{A}^{+}/\varepsilon_0^{+}$ [see SM~\cite{SM} for details].

The high tunability of exchange interactions in 2D systems \cite{martinez-carracedo_tuning_2024} allows precise control of the DW width down to the atomic scale. We fix $J_1=8$ meV, $J_3=1$ meV and $D=0.5$ meV, and vary $J_2$ toward the limit $6J_2\rightarrow-(J_1+4J_3)$, below which frustration destabilizes the system. As shown in Fig.~\ref{fig:fig4}(b) for three characteristic values of $J_2$, decreasing $J_2$ and reversing its sign flattens the low-energy branch ($\mathcal{A}^-\rightarrow0$), signaling instability. The curvature of the high energy mode eventually changes its sign, with $\mathcal{A}^{+}$ becoming negative, thereby modifying its contribution to the DW width. We obtain an accurate agreement with ASD simulations across the whole range of $J_2$ values [see Fig. \ref{fig:fig4}(c)].  

\begin{figure}[t]
    \centering
\includegraphics[width=0.465\textwidth]{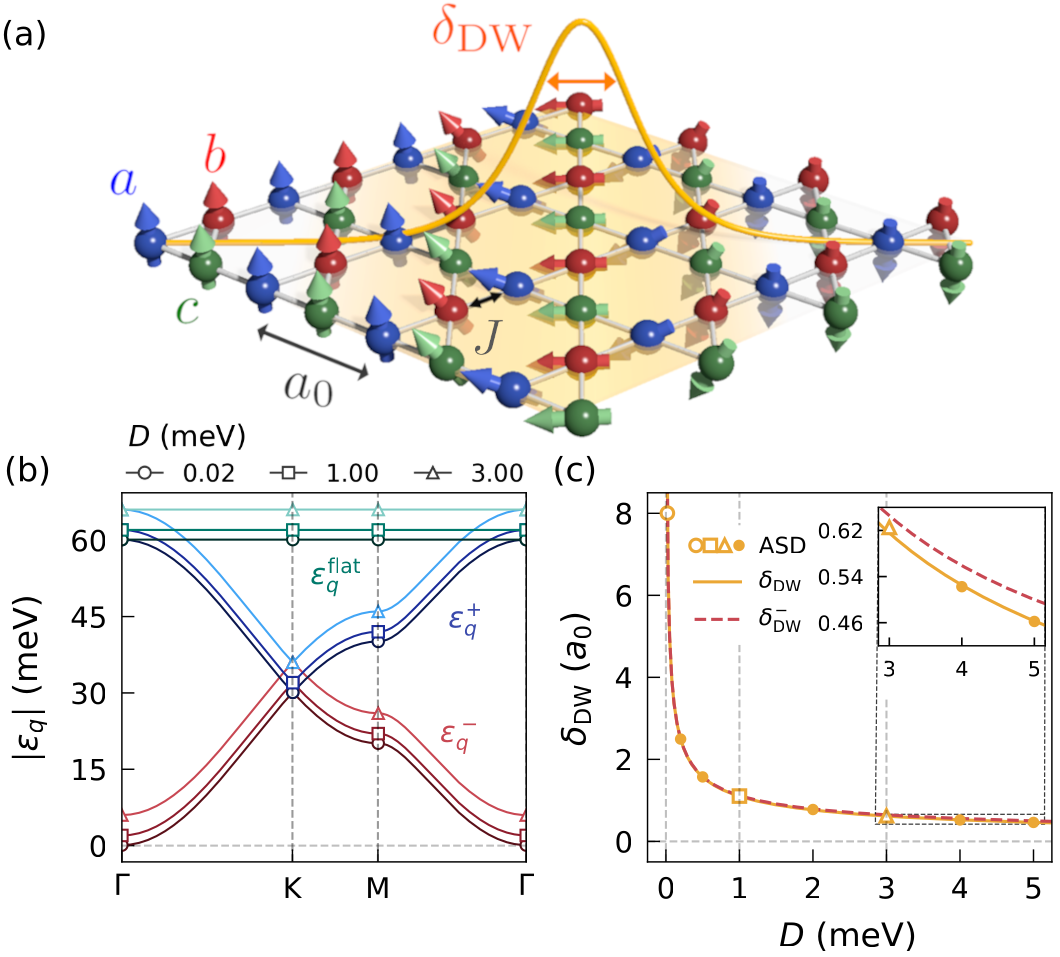}\caption{ 
(a) Sketch of a DW profile in a 2D kagome ferromagnet. (b) Spin-wave spectrum along the indicated high-symmetry path at zero temperature for selected values of the uniaxial anisotropy $D$. Solid lines denote LSWT results and symbols serve as a visual guide. (c) $\delta_{\rm DW}$ and $\delta_{\rm DW}^{-}$ contribution as a function of $D$. Solid lines correspond to analytical results and symbols correspond to ASD simulations, open symbols correspond to the selected cases shown in (b). The inset shows a zoom of the region with high anisotropies.}
\label{fig:fig5}
\end{figure}

As a final demonstration of the universality of our framework, we consider a 2D ferromagnetic kagome lattice composed of three sublattices [see Fig.~\ref{fig:fig5}(a)]. Kagome magnets host exotic quantum states \cite{han_fractionalized_2012} and are promising platforms for spintronics: compounds such as Mn$_3$Sn enable low-energy devices \cite{goren_antiferromagnetic_2025}, while materials like FeSn$_2$ exhibit a strong interplay between magnetism and nontrivial topological electronic properties \cite{ye_massive_2018}. Nevertheless, a fundamental understanding of the magnetic textures emerging in these systems remains incomplete \cite{li_discovery_2023}. We analyze a 2D kagome ferromagnet at zero temperature with fixed next-nearest-neighbor interaction \mbox{$J=10~\mathrm{meV}$} and increasing uniaxial anisotropy $D$.

The SW spectrum of the kagome FM lattice exhibits three branches [see Fig.~\ref{fig:fig5}(b)]. The DW width is described by Eq.~\eqref{eq:UniversalDW} as
\begin{equation}
\delta_{\rm DW}=\sqrt{\dfrac{J}{8D} -\dfrac{J}{8D+24J}},
\label{eq:kagome-contributions}
\end{equation}
where the terms under the square root correspond to the contributions from $\varepsilon_q^{\pm}$ modes respectively [see SM~\cite{SM} for details]. The flat band $\varepsilon_q^{\rm flat}$ does not contribute to the DW width expression, since $\mathcal{A}^{\mathrm{flat}} \propto \partial_q^2 \varepsilon_q\big|_{q=0}=0$.

As we increase the uniaxial anisotropy $D$, the energy gaps of the branches widen. Consequently, the second term in Eq.~\eqref{eq:kagome-contributions} becomes a more significant correction to the dominant contribution from the low-energy mode, $\varepsilon_q^{-}$. This correction enables a highly accurate estimation of the DW width up to relatively high anisotropies, where excellent agreement with ASD simulations is observed [see Fig.~\ref{fig:fig5}(c)].

In summary, we establish a universal microscopic framework that links the parameters defining the spin-wave spectrum of multi-sublattice Heisenberg magnets to the domain-wall width. This yields an analytical expression applicable to ferro-, antiferro- and ferrimagnets, as well as two-dimensional magnets. Our framework remains accurate from zero temperature up to the ordering transition. By incorporating thermal renormalization of microscopic couplings and validating the theory with large-scale atomistic simulations, we provide a unified description of domain-wall structure across diverse magnetic materials. Our results bridge atomistic spin interactions and continuum micromagnetics, offering a practical framework to interpret and engineer magnetic textures in complex spin systems.

\section{Acknowledgements}

J. M. L. and T. G. acknowledge financial support from PhD grants from Comunidad de Madrid (Grant No. PIPF-2023/TEC-29997 and No. PIPF-2021/TEC-25377). 
L. R. gratefully acknowledges funding from the National Research, Development, and Innovation Office (NRDI) of Hungary under Project Nos. FK142601 and ADVANCED 149745, as well as from the Hungarian Academy of Sciences via a János Bolyai Research Grant (No. BO/00178/23/11). 
U. A. acknowledge support from PIE grant from CSIC
(20226AT018).
We acknowledge support from grants PID2024-157112OB-C52, CNS2023-144681 and CEX2024-001445-S funded by MCIN/AEI/10.13039/501100011033 and by “ERDF A way of making Europe” and “ESF Investing in your future.” 
R. M. O. and U. A. are grateful for financial support from COST Action CA23136 CHIROMAG.

\bibliography{Ref}

\end{document}